\documentclass[12pt,preprint]{aastex}
\def\psr{PSR~J1740$-$5340 }
\def\com{COM~J1740$-$5340 }

\def\ltsima{$\; \buildrel < \over \sim \;$}
\def\gtsima{$\; \buildrel > \over \sim \;$}
\def\lsim{\lower.5ex\hbox{\ltsima}}
\def\gsim{\lower.5ex\hbox{\gtsima}}
\def\lapp{\ifmmode\stackrel{<}{_{\sim}}\else$\stackrel{<}{_{\sim}}$\fi}
\def\gapp{\ifmmode\stackrel{>}{_{\sim}}\else$\stackrel{<}{_{\sim}}$\fi}

\newdimen\minuswidth    
\setbox0=\hbox{$-$}
\minuswidth=\wd0
\catcode`@=\active
\def@{\kern\minuswidth}
\newdimen\digitwidth    
\setbox0=\hbox{\rm0}
\digitwidth=\wd0
\catcode`!=\active
\def!{\kern\digitwidth}
 
\begin{document} 
  
\title{Accurate mass ratio and heating effects in the
      dual-line millisecond binary pulsar in NGC 6397
\footnote{Based on observations collected at the
European Southern Observatory, Chile, proposal 69.D-0264}
}

\author{F.R. Ferraro\altaffilmark{2}, 
E. Sabbi\footnote{Dipartimento di Astronomia Universit\`a 
di Bologna, via Ranzani 1, I--40127 Bologna, Italy,
ferraro@bo.astro.it, sabbi@bo.astro.it},
R. Gratton\footnote{INAF--Osservatorio Astronomico di Padova,
vicolo dell'Osservatorio 5, I--35122 Padova, Italy, gratton@pd.astro.it},
A. Possenti\footnote{INAF-Osservatorio Astronomico di Cagliari,
Loc. Poggio dei Pini, Strada 54, I--09012 Capoterra, Italy}, 
N. D'Amico\footnote{INAF- Osservatorio Astronomico di Cagliari,
Loc. Poggio dei Pini, Strada 54, I--09012 Capoterra, Italy
\&   Dipartimento di Fisica Universit\`a di Cagliari,
Cittadella Universitaria, I-09042 Monserrato, Italy,
damico@ca.astro.it},
A. Bragaglia\footnote{INAF -  Osservatorio Astronomico
 di Bologna, via Ranzani 1, I--40127 Bologna, Italy, angela@bo.astro.it},
F. Camilo\footnote{Columbia Astrophysics Laboratory, Columbia
University, 550 West 120th Street, New York, NY 10027,
fernando@astro.columbia.edu}
}
\medskip

\begin{abstract}
By means of high-resolution spectra 
we have measured radial velocities of the companion
(hereafter \com$\!\!$) to the eclipsing millisecond pulsar \psr in the
Galactic globular cluster NGC 6397.  The radial-velocity curve fully
confirms that \com is orbiting the pulsar and enables us to derive the
most accurate mass ratio ($M_{{\rm PSR}}/M_{\rm {COM}}=5.85\pm0.13$)
for any non-relativistic binary system containing a neutron star.
Assuming a pulsar mass in the range $1.3-1.9~{\rm M_\odot}$, the mass
of \com spans the interval $0.22-0.32~{\rm M_{\odot}}$, the
inclination of the system is constrained within $56^{\circ}\gapp
i\gapp 47^{\circ}$ and the Roche lobe radius is $r_{RL}\sim
1.5-1.7~{\rm R_{\odot}}.$ A preliminary chemical abundance analysis
confirms that \com has a metallicity compatible with that measured for
other stars in this metal-poor globular, but the unexpected detection
of strong He{\sc i} absorption lines implies the existence of regions
at $T>10,000$\,K, significantly warmer than the rest of the star. The
intensity of this line correlates with the orbital phase, suggesting
the presence of a region on the companion surface, heated by the
millisecond pulsar flux.
\end{abstract}

\keywords{Globular clusters: individual (NGC~6397) --- stars:
evolution --- binaries: close --- pulsars: individual (PSR
J1740$-$5340) --- techniques: spectroscopic}

\section{Introduction} 
\label{intro}
During a systematic search of the galactic globular cluster (GC)
system for millisecond pulsars (MSPs) carried out with the Parkes
radio telescope, D'Amico et al. (2001a) discovered the binary
millisecond pulsar \psr, in the nearby globular cluster NGC 6397. The
pulsar displays eclipses at a frequency of 1.4 GHz for more than 40\%
of the $\sim 32.5$ hr orbital period and exhibits striking
irregularities of the radio signal at all orbital phases. These
features indicate that the MSP is orbiting within a large envelope of
matter released from the companion (D'Amico et al. 2001b), whose
interaction with the pulsar wind could be responsible for the
modulated and probably extended X-ray emission detected with {\it
Chandra} (Grindlay et al. 2001, 2002).  By using the position of the
MSP inferred from radio timing, Ferraro et al. (2001) identified a
variable star ({\it Star A}, hereafter COM~J1740$-$5340) 
whose optical modulation
nicely agrees with the orbital period of the MSP.  In particular,
Ferraro et al. noticed that \com is the first example of a MSP
companion whose light curve is dominated by ellipsoidal variations,
suggestive of a tidally distorted star, which almost completely fills
(and is still overflowing) its Roche lobe.

Binary evolution calculations (e.g. Tauris \& Savonije 1999;
Podsiadlowski, Rappaport \& Pfahl 2002) and the few optical detections
(e.g. Hansen \& Phinney 1998; Stappers et al. 2001) show that the
common companion to a binary MSP is either a white dwarf or a very
light ($\sim 0.01-0.03~{\rm M_\odot}$) almost completely exhausted
(and perhaps evaporating) star. If a MSP is located in a GC, dynamical
encounters in the cluster core can also provide it with other kinds of
companion and in fact a main sequence star has been probably
identified as the star orbiting the MSP 47Tuc-W in 47 Tucanae (Edmonds
et al. 2002). None of these hypotheses can be applied to \com$\!\!:$
it is too luminous to be a white dwarf (V$\sim 16.5$, comparable to
the turn-off stars of NGC~6397); its mass ($M_{{\rm COM}}\ge 0.19~{\rm
M_\odot}$ for a $1.4~{\rm M_\odot}$ neutron star, D'Amico et
al. 2001a) is not compatible with that of a very light stellar
companion; its anomalous red color would require it to be perturbed if
it was originally on the main sequence. 
As a consequence, a wealth of intriguing
 scenarios
have flourished in order to explain the nature of the binary (see
Possenti 2002, Orosz \& van Kerkwijk 2003, Grindlay et al. 2002 for a review).

In any model, \psr must have undergone at least one and perhaps
multiple phases of ``recycling'' (Alpar et al. 1982).  During this
process significant accretion of mass ($\sim 0.1-0.7~{\rm {M_\odot}}$)
onto the neutron star (NS) is expected (van den Heuvel \& Bitzaraki
1995; Burderi et al. 1999), but up to now no observation has measured
a MSP NS mass unambiguously outside the narrow range of masses
(average $1.35\pm 0.04~{\rm {M_\odot}}$) measured by Thorsett \&
Chakrabarty (1999) in a sample of binary pulsars statistically
dominated by NS-NS relativistic systems.  Besides clarifying the
spin-up mechanism, finding a more massive NS could shed light also on
the equation of state for nuclear matter (e.g. Cook, Shapiro \&
Teukolsky 1994).

Interestingly, \psr and its companion constitute the optically
brightest dual-line binary hosting a MSP, thus being a prime target
for an accurate determination of both the binary parameters and the
mass of a spun-up pulsar, while yielding the possibility of
discriminating between the different suggested evolutionary paths.
Hence, we have planned a coordinated spectro-photometric campaign
using the ESO telescopes in La Silla and Paranal (both in Chile).
While the photometric survey is still in progress, we present in this
{\it Letter} the first spectroscopic results of the project, based on
the analysis of a set of phase-resolved high-resolution spectra. Here
we present the radial velocity curve of COM~J1740$-$5340, the mass 
ratio of the
system and preliminary estimates of its metallicity and effective
temperature. The detailed analysis of the chemical abundance pattern
of \com will be presented in a forthcoming  paper (Sabbi et al 2003, 
in preparation).
 
\section{Observations and  data analysis}
\label{obs}

The observations were performed in service mode with the {\it
Ultraviolet-Visual Echelle Spectrograph} (UVES) mounted at the {\it
Kueyen} 8m-telescope (UT2) of the ESO Very Large Telescope on Cerro
Paranal (Chile).  The spectra were obtained on 8 different nights from
2002 May to 2002 June (see Table \ref{t:mjd}) in order to cover the
complete orbital period of the system.  The dichroic beamsplitter \#1
together with grisms \#2 (centered at $3900~$\AA) and \#3 (centered at
$5800~$\AA) were used: this configuration allowed us to observe
simultaneously two spectral ranges with the two arms (blue and red) of
the spectrograph, covering the wavelength ranges $3280-4490~$\AA,
$4725-5708~$\AA, and $5817-6725~$\AA. A 1\arcsec ~wide and 8\arcsec
~long slit was adopted, yielding a resolution of $R\sim 40,000$ at
order centers.  The exposure time for each spectrum ($2600$ sec, equal
to $\sim 2$\% of the system orbital period) enabled us to perform
accurate phase resolved analysis, with a typical signal-to-noise ratio
$S/N\ge 20$ per pixel, measured at the continuum level.  All the
spectra were extracted with the UVES pipeline (Ballester et al. 2000).

\section{Results and Discussion}
\subsection{Radial Velocity Curve}  
\label{Vrad}
The broad spectral range covered by the spectra allows the observation
of a large number of spectral features.  In particular we selected a
set of $\sim 10-20$ lines (comprising Fe{\sc i}, Ca{\sc i}, Na{\sc i},
Mg{\sc i}, Mn{\sc i}, Ti{\sc i} and H$_{\gamma}$) not contaminated by
atmospheric and/or interstellar contributions.  The Doppler-shifted
wavelength of each line has been measured: the resolution of the UVES
spectra allows very accurate measures of line centroids, with a
typical formal accuracy of $\sim 0.02-0.05$ \AA ($\sim 1-3$ km
s$^{-1}$).  We have verified that the broadening of each line due to
the variation of the velocity of the source along a 2600-sec
integration is always negligible with respect to the intrinsic width
of the lines.  The line broadening due to rotation ($V_{\rm
rot}=49.6\pm 0.9$ km s$^{-1}$) permitted
rejection of contaminating lines coming from nearby, non rotating
objects (the system is in a crowded region, see Fig.~1 of Ferraro et
al. 2001).  The wavelength of each line has been converted to radial
velocity (RV),  taking into account for the heliocentric 
correction. In order to
minimize spurious effects we measured RVs independently on the red and
blue part of each spectrum and carefully checked that no significant
offset was present.  Then, all the RV measures obtained in each
spectrum were averaged and a mean value was obtained.

\begin{figure} 
\plotone{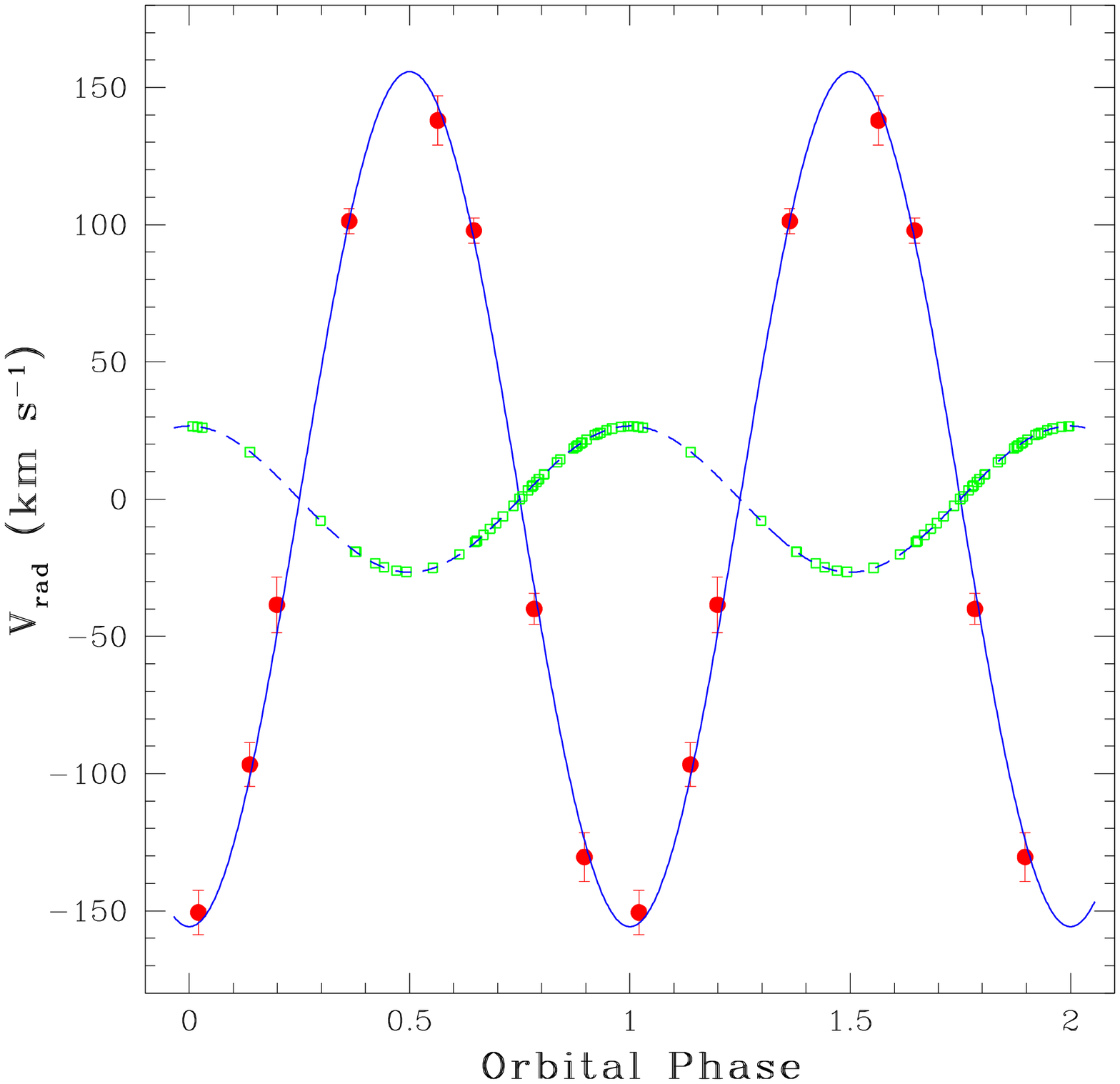} 
\caption {
The {\it large dots} are the radial velocity determinations for \com. The
{\it solid} line represents the best-fit sinusoidal curve. 
The {\it small open squares} are the radial velocity determinations for
\psr derived from timing measurements and the radio ephemeris (D'Amico et
al. 2001b).  The {\it dashed} line represents the fitted velocity
curve of the pulsar.\label{Vcurve} }
\end{figure}

In order to determine the amplitude $K$ of the RV curve and the
systemic velocity $\gamma$, we fitted the data using the sum of a
constant and a sinusoidal function, which is adequate to describe the
orbital motion of this almost circular system (eccentricity
$e<10^{-4}$; D'Amico et al. 2001b).  The best fit yields a
radial-velocity amplitude $K=155.8 \pm 3.6$ km s$^{-1}$
($1\sigma$-error) with $\chi^2_{\nu}=0.99$.  Inspection of Figure
\ref{Vcurve} shows that the absolute phase of the RV data matches that
of the radio pulsar orbital motion very well (the radial velocities
are null at inferior and superior conjunctions which, according to the
convention of D'Amico et al. 2001b, correspond to orbital phases 0.25
and 0.75 respectively), confirming unambiguously that \psr and \com
orbit each other.  The systemic velocity is $\gamma =17.7\pm 2.3$ km
s$^{-1}$, which is in perfect agreement with the radial motion of the
cluster ($V_{{\rm 6397}}=18.9\pm 0.1$ km s$^{-1}$; Harris 1996),
further confirming the membership
\footnote{The small value of the difference $|\gamma - V_{{\rm
6397}}|<3.6$ km s$^{-1}$ ($1\sigma$ limit) suggests that the center of
mass of the binary is now near apoastron of a highly elliptical orbit
in the globular cluster.  In fact, were the binary on an almost
circular orbit at 11 core radii from the GC center, its relative line
of sight velocity (estimated from the enclosed mass) would be of the
order $\gapp 10$ km s$^{-1}.$ This result supports the hypothesis that
the binary has been recently kicked out of the core of NGC 6397 due
the a dynamical interaction}. Table \ref{t:mjd} lists the final
heliocentric velocities after subtraction of the $\gamma$ velocity,
where the standard deviation of the average has been assumed as an
estimate of the error.  The epoch and the orbital phase of the
mid-time of each observation are also reported.  The latter has been
computed using the radio ephemeris of \psr (D'Amico et al. 2001b).
The resulting radial velocity-curve is shown in Figure~\ref{Vcurve}.

\subsection{Constraints on the mass of \com}  
\label{Mass}

\begin{figure} 
\plotone{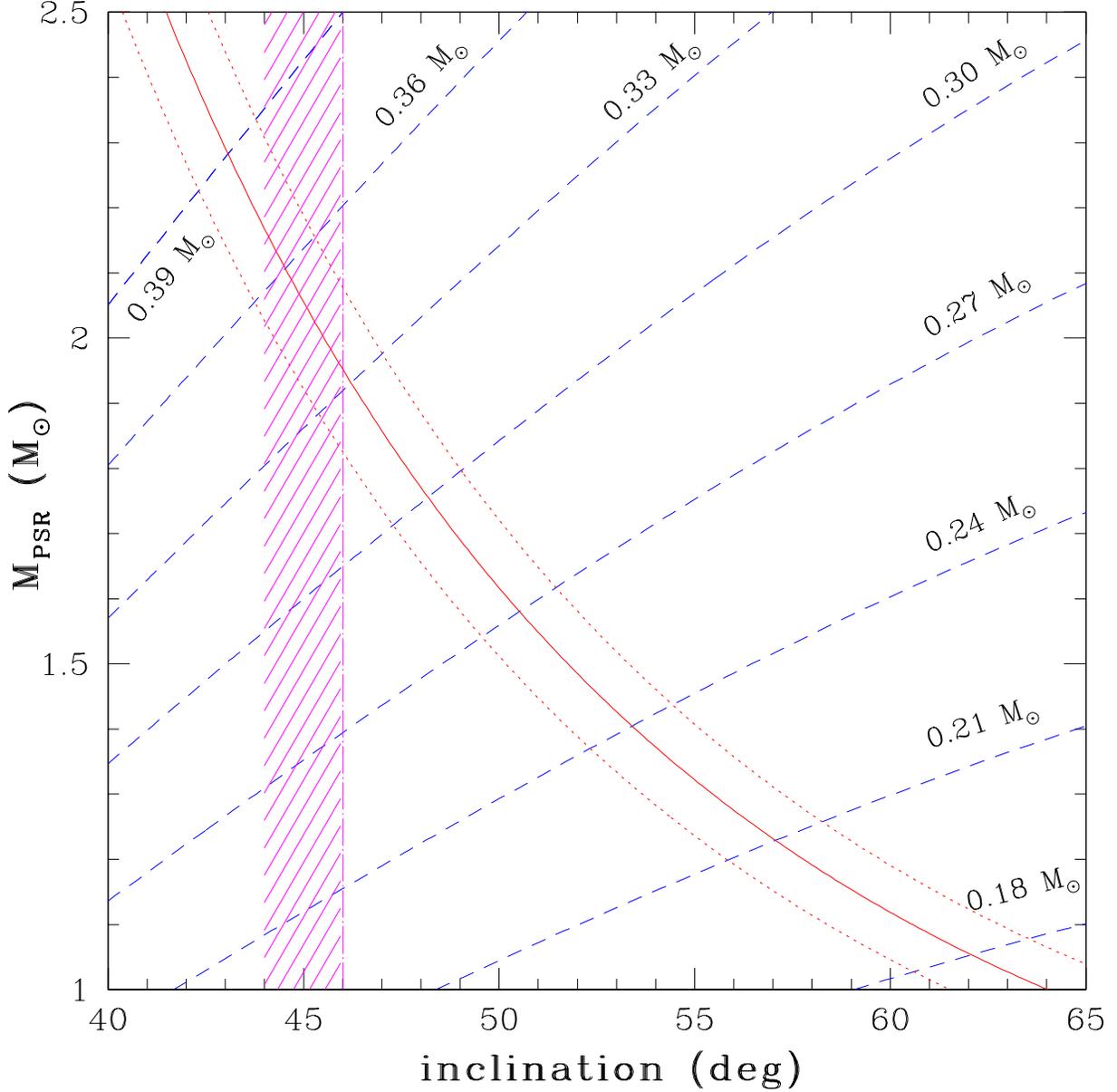} 
\caption {\label{MvsM} 
Mass of \psr and orbital inclination of the binary. The allowed range
of values are constrained to lie within the narrow strip whose borders
({\it dotted lines}) are the $1\sigma$ boundaries derived from the
mass ratio of the 
 system. Lines of constant mass for \com are also
shown ({\it dashed lines}) and labeled with the assumed mass
value. The vertical {\it dot-dashed line} represents the $2\sigma$
lower limit on the orbital inclination derived from the modeling of
the light curves of \com (Orosz \& van Kerkwijk 2003). This lower
limit is nearly coincident with that inferred from the rotational
velocity of the companion.}
\end{figure}

Being a dual-line binary pulsar with a very bright companion, this
system can be used to infer tight constraints to the masses of the
components. The mass function $f(M_{\rm COM})$ of \com can be easily
computed from the radial velocity amplitude $K$ and the orbital
period $P_{\rm orb}=1.35405939\pm (5\times 10^{-8})$ days (D'Amico et
al. 2001b):
\begin{equation}
f(M_{\rm COM}) = {{M_{\rm PSR}^{3} \sin^{3}i}\over
{(M_{\rm COM}+M_{\rm PSR})^{2}}}
= {{K^3 P_{\rm orb}}\over{2\pi G}}=0.530 \pm 0.038~{\rm M_{\odot}}~,
\label{eq:mfcom}
\end{equation}
where $M_{\rm PSR}$ and $M_{\rm COM}$ are the masses of the pulsar and
the companion star, $i$ is the inclination of the orbital plane
with respect to the line-of-sight
and the errors are quoted at the $1\sigma$-level, as
everywhere in the following. The mass ratio $q=M_{\rm PSR}/M_{\rm
COM}$ can be derived by combining eq. (\ref{eq:mfcom}) with the mass
function of the MSP obtained by D'Amico et al. (2001b):
\begin{equation}
f(M_{\rm PSR}) = {{M_{\rm COM}^{3} \sin^{3}i}\over
{(M_{\rm COM}+M_{\rm PSR})^{2}}}
=(2.6442\pm 0.0003)\times 10^{-3}~{\rm M_{\odot}}~.
\label{eq:mfpsr}
\end{equation}
We obtain $q=5.85\pm 0.13$, which is the most accurate estimate ($\sim
2$\% error) ever obtained for the mass ratio in any non-relativistic
binary comprising a neutron star and in particular a recycled pulsar.
In order to completely solve for the binary parameters one needs the
orbital inclination of the system. However, for a reasonable choice of
the neutron star mass (in the range $1.0-2.5~{\rm M_{\odot}}$, Shapiro
\& Teukolsky 1983), the precise measurement of the mass ratio reduces
the allowed space of the parameters to a very narrow strip in the
$M_{\rm PSR}$ vs. $i$ diagram (see Fig.~\ref{MvsM}).

A further constraint on the inclination can be inferred from the
ellipsoidal modulation of the light curves of \com (the ongoing
photometric analysis will be presented elsewhere).  Preliminary
results, based on an incomplete coverage of the orbit (Orosz \& van
Kerkwijk 2003) indicate $i>46^{\circ}$ (at $2\sigma$) which implies
(Fig. \ref{MvsM}) $M_{\rm PSR}\lapp 2.0$.  This lower limit on $i$
well agrees with that ($i>46^{\circ}.3$) derived from our determination
of the rotational velocity $V_{\rm rot}$ of \com assuming a
photometric radius of the companion in the range ($1.60\pm 0.17~{\rm
R_\odot}$) measured by Orosz \& van Kerkwijk (2003).

Hence in Table
\ref{t:par}, we have listed some of the orbital parameters of the
system derived assuming the three reference cases of $M_{\rm
PSR}=1.3,1.5,1.9~{\rm M_{\odot}}$.  In all these cases, the radius of
the Roche lobe of \com matches with the dimension estimated from HST
photometry (Ferraro et al. 2001). The implied range of the companion
masses ($0.22~{\rm {M_\odot}}\lapp M_{\rm COM}\lapp 0.32~{\rm
{M_\odot}}$) fits all the evolutionary models proposed so far
(Burderi, D'Antona \& Burgay 2002; Orosz \& van Kerwijk 2003; Grindlay
et. al 2002; Ergma \& Sarna 2003), whose discrimination will be
 greatly assisted by  the detailed measurement of the
chemical abundances (Sabbi et al 2003, in preparation).

\subsection{Chemical composition analysis: a surprising result}
\label{chemcomp}

\begin{figure} 
\plotone{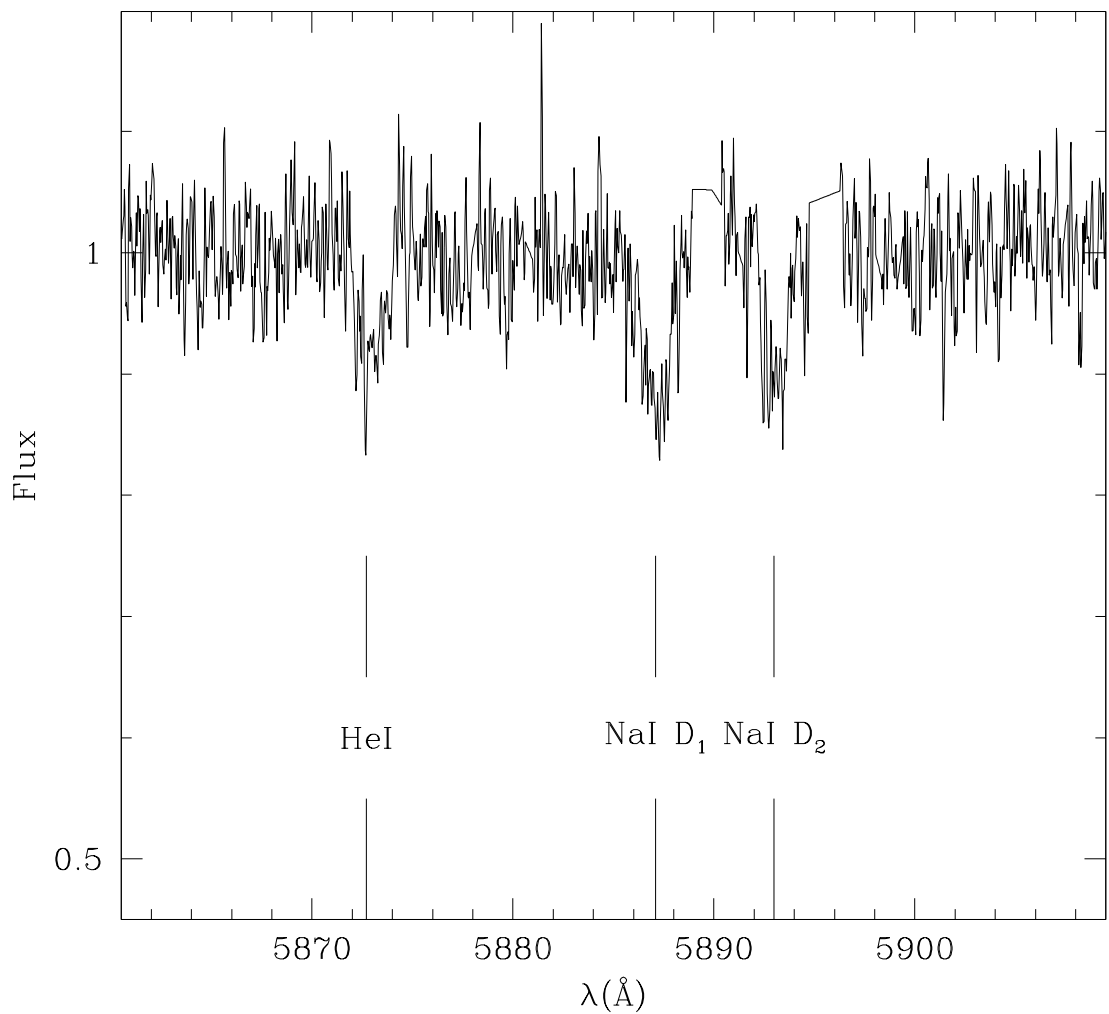} 
\caption{\label{He} 
Portion of the normalized spectrum taken at $\phi=0.02$ in the Na{\sc
i} D lines region (the telluric Na{\sc i} lines have been removed for
clarity).  The spectrum has been smoothed with a boxcar of 3 pixels,
and not shifted to rest wavelengths; the observed RV is $-145.4$ km
s$^{-1}$.  The He{\sc i} absorption line at 5875.6 \AA ~is clearly
visible; Na{\sc i} D lines and the strong He line are indicated.  }
\end{figure}

In order to perform a preliminary chemical abundance analysis, the two
spectra taken under the best seeing conditions (at phases 0.021 and
0.564, close to the quadratures, see Table \ref{t:mjd}) have been
corrected for the RV and then combined, thus attaining $S/N\sim 45$ in
the wavelength region around 5900 ~\AA, near the Na{\sc i} D lines. 
The estimate of the metallicity has been
obtained inspecting the equivalent width of about 40 useful lines
(mostly Fe{\sc i}) and by assuming a gravity (log $g=3.2$) compatible
with the position of \com in the color-magnitude diagram.
Figure \ref{He} shows a zoom-in on that spectral 
region of the normalized spectrum 
taken at phase 0.02.  

In summary, the main features of the spectra are: (1) strong
broadening of the lines due to the rotation of the star; (2) 
a low value of the metallicity ${\rm [Fe/H]}\sim -2$ and an
excess of $\alpha$-elements such as Na{\sc i}, Mg{\sc i}, and Si{\sc i};
(3) the presence of many lines of low excitation, together with a few weak
lines of ionized elements.
Feature (2) is fully compatible with the results derived for other
stars in this cluster (Gratton et al. 2001), and further 
confirms the membership of the object to NGC 6397.
Feature (3) indicates that the effective temperature is
low, in agreement with the red color of the source.  The ${\rm
H}_{\alpha}$ line wings imply $T_{{\rm eff}}\simeq 5530\pm 70$ K,
confirming the previous estimate of Ferraro et al. (2001).
  
The most fascinating result of the preliminary abundance analysis is
the presence of He{\sc i} absorption lines at 5875.6 \AA~ and at
6678.2 \AA. The former line is clearly visible in Figure
\ref{He}. Such a spectral feature is completely unexpected in a
low-temperature star like COM~J1740$-$5340, being the signature of photospheric
regions at $T>10,000$\,K.  Remarkably, the He{\sc i} lines are clearly
visible in all the spectra, with the exception of the two taken at
phases 0.199 and 0.363, when \com is located between the MSP and the
observer.  This result suggests the existence of a region of the star
(facing the pulsar) at a temperature significantly larger than the
rest of the surface. More careful analyses are under way to
assess the dimensions of the heated region.  However, the relatively
small orbital variation of the colors (hence effective temperature) of
the star (Orosz \& van Kerkwijk 2003, Kaluzny, Rucinski, \& Thompson. 2002)
is suggestive of a heated area, smaller than the cross
section of the surface of \com visible from the MSP.  At the same
time, the rotation profile of the He{\sc i} lines would  suggest
that this warmed region extends rather longitudinally along
the companion surface (an equatorial strip?).  As an example, a region
facing the pulsar with a dimension $\sim 3-5$\% of the Roche lobe
radius of \com and with $T\sim 10^4$ K, could produce a brightening of
0.02--0.04 mag in the light curves of \com near orbital phase 0.75
(when we see the hemisphere of the companion facing the pulsar). This
could explain the asymmetry in the light curves seen around the
minimum at phase 0.75 (Ferraro et al. 2001).

The existence of a strongly heated portion of the surface of a MSP
companion is not uncommon (e.g. Stappers et al. 2001) and naturally
calls for the effects of the energetic flux of particles and
electromagnetic waves released from the MSP. What is peculiar in \com
is the strip-like shape and the equatorial location of the heated
region.  This would suggest that the pulsar energetic flux is
preferentially emitted in the orbital plane of the binary with a
highly flattened shape.  Noticeably, a similar planar emission pattern
of the pulsar wind has been invoked in order to explain the origin of
the X-ray inner ring surrounding the Crab, whose spin axis is believed to be
nearly orthogonal to the plane containing the ring (Hester et al. 2002).

In summary, if confirmed by future analysis, the existence of a warmed
strip onto the surface of \com would be of great significance: first,
a strongly anisotropic emission could enhance the amount of the pulsar
rotational energy impinging onto the companion. This could trigger yet
largely unmodeled effects during the radiation-driven evolution of a
MSP binary system (D'Antona 1996).  Secondly, location and
extension of the heated strip could be used for inferring the MSP spin
and magnetic axes orientations, helping in discriminating among
the different evolutionary scenarios proposed so far.
 
\acknowledgements{\small Financial support for this research has been
provided by the Agenzia Spaziale Italiana (ASI). FC is
supported by NASA grant NAG~5-9950. We thank the referee J.E. Grindlay for his
precious suggestions.}

\clearpage

\clearpage

\clearpage

\clearpage

\begin{deluxetable}{lcr}
\scriptsize \tablewidth{10cm}  
\tablecaption{\label{t:mjd}
Radial velocity measurements}
\startdata \\
\hline\hline
MJD & Orbital Phase & Radial Velocity   \\
    &               & (km s$^{-1}$)    \\
\hline 
52413.22761 & 0.0204 & $-150.6\pm8.0$    \\
52405.26219 & 0.1374 & $ -96.7\pm8.0$     \\
52416.17728 & 0.1988 & $ -36.7\pm10.1$    \\
52404.21301 & 0.3630 & $ 101.3\pm4.6$    \\
52434.27455 & 0.5640 & $ 138.0\pm8.9$    \\
52422.19896 & 0.6460 & $  97.9\pm4.5$     \\
52414.26020 & 0.7830 & $ -40.0\pm5.7$     \\
52440.14142 & 0.8968 & $-130.4\pm8.9$     \\
\hline
\enddata
\end{deluxetable}

\begin{deluxetable}{lccc}
\scriptsize \tablewidth{17cm} 
\tablecaption{\label{t:par} 
Derived parameters of the \psr system}
\startdata \\
\hline\hline
Mass ratio,  $q$ &  & 5.85 $\pm$ 0.13 & \\
Temperature of \com, $T_{\rm eff}$ (K) &  & $5530\pm 70$  \\
Radial-velocity amplitude of \com, $K$ (km s$^{-1}$) &  & $155.8\pm 3.6$ & \\ 
\hline
Mass of \psr, ($M_{\odot}$)                  & 1.30 & 1.50 & 1.90 \\
Mass of \com, ($M_{\odot}$)                  & 0.22 & 0.26 & 0.32 \\
Inclination angle, $i$ (deg)                 & 56   & 51   &  47  \\
Orbital separation, $a$ ($R_{\odot}$)        & 6.1  & 6.5  & 7.0  \\
Roche lobe radius of \com, $r_{RL}$ ($R_{\odot}$) & 1.5  & 1.6  & 1.7  \\
\hline
\enddata
\end{deluxetable}

\end{document}